\documentstyle[sprocl,epsfig]{article}

\def\be{\begin{equation}}
\def\ee{\end{equation}}
\def\bea{\begin{eqnarray}}
\def\eea{\end{eqnarray}}

\begin{document}

\title{BARYON NUMBER ASYMMETRY INDUCED BY COHERENT MOTIONS OF A
       COSMOLOGICAL AXION-LIKE PSEUDOSCALAR}

\author{RAM BRUSTEIN$\footnote{Based on a talk at COSMO 99: 3rd International Conference on Particle \\ 
Physics and the Early Universe, Trieste, Italy, 27 Sep - 3
Oct 1999.}$ and DAVID H. OAKNIN}

\address{Department of Physics, Ben-Gurion University, \\
Beer-sheva 84105, Israel\\E-mail: ramyb,doaknin@bgumail.bgu.ac.il}

\maketitle\abstracts{ A cosmological pseudoscalar field coupled to
hypercharge topological number density can exponentially amplify
hyperelectric and hypermagnetic fields in the symmetric phase of
the electroweak plasma while coherently rolling or oscillating, leading to
the formation of a condensate of topological number density. The
topological number can be converted, under certain conditions,
into  baryon number in sufficient amount to explain the observed
baryon asymmetry of the universe. We focus on a singlet elementary
hypercharge axion (HCA) whose only coupling to standard model fields is to
hypercharge topological number density. This model has two
parameters: the mass $m$ and decay constant $M$ of the HCA. We
describe this new mechanism for baryogenesis and outline the
region of the parameter space in which the mechanism is efficient.
We show that present colliders can already put interesting
constraints on both parameters, and that future colliders will
improve the detection capabilities very significatively.}

\section{Introduction.}
The origin of the baryon asymmetry of the universe  remains
one of the most fundamental open questions in high energy physics
and cosmology. In 1967 Sakharov noticed \cite{Sakharov:1967dj}
that three conditions are essential for the creation of a net
baryon number in a previously symmetric universe: 1) baryon number
non-conservation; 2) C and CP violation; 3) out of equilibrium
dynamics. Since then many different hypothetical scenarios for
baryogenesis have been proposed. A dramatic conclusion  emerged
from the studies of these scenarios: new physics beyond the
Standard Model (SM) is required to explain baryogenesis
\cite{Shaposhnikov:1986jp}.

It has been recently realized 
\cite{Giovannini:1998eg,Elmfors:1998wz,Grasso:1999me} that
topologically non-trivial configurations of hypercharge gauge
fields can be relevant players in the electroweak (EW) scenario for
baryogenesis. Hypercharge fields  couple anomalously to fermionic
number densities in the symmetric phase of the EW plasma, while
their surviving long-range projections onto usual electromagnetic
fields in the broken phase of the plasma do not. As a consequence,
the hypercharge Chern-Simons (CS) number stored in the symmetric
phase just before the transition can be converted into a fermionic
asymmetry along the direction $B-L=0$ when the EW symmetry is
spontaneously broken.

A hypothetical axion-like pseudoscalar field coupled to
hypercharge topological number density can amplify hyperelectric
and hypermagnetic fields in the unbroken phase of the EW plasma,
while coherently rolling or oscillating around the minimum of its
potential. The coherent motion provides the three Sakharov's
conditions and is capable of generating a net CS number that can
survive until the phase transition and then be converted into
baryonic asymmetry \cite{Brustein:1999du,Brustein:1999rk}, as first noted
in [8]. The
mechanism could explain the origin of the baryon number of the
universe, if the EW phase transition is strong enough such that
the generated asymmetry is not erased by B-violating processes in
thermal equilibrium in the broken phase of the plasma.

Pseudoscalar fields with axion-like coupling to CS number
densities appear in several possible extensions of the SM. They
were originally proposed as an elegant solution to the strong
CP-problem. In models with an extended higgs-sector the physical
pseudoscalar can couple to hypercharge topological density through
quantum effects. In supergravity or superstring models axions that
couple to extended gauge groups are common.

Experimental signatures of this hypothetical particle could appear
in present and/or future colliders. Since the hypercharge photon
is a linear combination of the ordinary photon and the $Z$, the
particle can be produced in association with a photon or a $Z$,
and detected through its decay into a pair of neutral gauge
bosons, if these signatures are not overshadowed by their SM
backgrounds \cite{Brustein:1999it}.

\section{The model}

We will assume that  the  universe is  homogeneous and isotropic,
and can be described by a conformally flat metric. In addition to
the SM fields we consider a  time-dependent pseudoscalar field
$\phi(\eta)$ with coupling $\frac{\lambda}{4} \phi Y\widetilde{Y}$
to the $U(1)_Y$ hypercharge field strength and a potential $V_0^4
V(\phi/f)$ generated by processes at energies higher than the EW
scale. We will also assume that the universe is radiation
dominated at some early time before the scalar dynamics becomes
relevant. The coupling constant $\lambda=1/M$ has units of
mass${}^{-1}$. For a typical axion coupled to a non-abelian gauge
topological density the potential is generated by non-perturbative
effects at the confinement scale $V_0$, and $f\sim M$.  In
general, this is not always the case, and we will allow
$\Lambda\equiv f/M>1$, but keeping the pseudoscalar mass $m\sim
V_0^2/f$ much smaller than the scale $M$.

Maxwell's equations describing hyper EM fields in the symmetric
phase of the highly conducting EW plasma, coupled to the heavy
pseudoscalar are the following,
\begin{eqnarray}
&(i)&\ \nabla \cdot{\vec E} = 0 \hspace{.3in} (ii)\ \nabla \cdot
{\vec B} = 0 \cr &(iii)&\ {\vec J} = \sigma {\vec E}
\hspace{.4in} (iv)\ \frac{\partial {\vec B}}{\partial \eta} =
-\nabla \times {\vec E} \cr &(v)&\ \frac{\partial {\vec
E}}{\partial \eta} =  \nabla \times{\vec B} - \lambda \frac{d
\phi}{d \eta} {\vec B} - {\vec J}.\ \ \ \label{maxwell}
\end{eqnarray}
Equations $(i)$ and $(iii)$ are valid for wavelengths larger than
the typical collision length $r_{Debye} \sim T^{-1}$ in the hot
plasma, so that individual charges are screened by collective
effects. The description of short wavelength modes should take
into account charge separation in the medium.

We have rescaled the electric and magnetic fields $\vec
E=a^2(\eta) \vec{\cal E}$, $\vec B=a^2(\eta) \vec{\cal B}$ and the
physical conductivity $\sigma=a(\eta)\sigma_c$, where $a(\eta)$ is
the scale factor of the universe and $\eta$ is conformal time. In
the EW plasma $\sigma_c \sim 10 T$. The fields $\vec{\cal E}$,
$\vec{\cal B}$ are the flat space EM fields, and we have assumed
for simplicity vanishing  bulk velocity ${\vec v}$ of the plasma
and zero chemical potentials for all species.

The equation for the pseudoscalar $\phi$ is the following
\begin{equation} \label{axion} \frac{d^2 \phi}{d
\eta^2} + 2 a H \frac{d \phi}{d \eta} + a^2 m^2 \phi = \lambda a^2
{\vec E} \cdot{\vec B},
\end{equation}
where $H=\frac{1}{a^2} \frac{da}{d\eta}$ is the Hubble parameter.
We will neglect the backreaction of the electromagnetic fields on
the scalar field since is irrelevant for most of the physics we
would like to explore \cite{Brustein:1999rk}. We therefore solve
eq.({\ref {axion}}) with vanishing r.h.s., and substitute the
resulting $\phi(\eta)$ into eq.({\ref{maxwell}).

\section{Amplification of Primordial Hypermagnetic Fields}

We will describe solutions to eq.({\ref{maxwell}) of the form
$\vec E({\vec x},\eta)= \int d^3{\vec k}\ e^{-i{\vec k}{\vec x}}\
{\vec e_{\vec k}}\ \epsilon_{\vec k}(\eta)$, $\vec B({\vec
x},\eta)= \int d^3{\vec k}\ e^{-i{\vec k}{\vec x}}\ {\vec b_{\vec
k}}\ \beta_{\vec k}(\eta)$, for which the electric and magnetic
modes are parallel to each other.  We find that the Fourier modes
${\vec e}_{\vec k}$ and ${\vec b}_{\vec k}$ are related,
$
{{\vec b}_{\vec k}}{}^{\pm}= b_k^\pm  ({\hat e}_1 \pm i{\hat
e}_2), \ \ \ k {{\vec e}_{\vec k}}{}^{\pm} \epsilon^\pm_{\vec
k}(\eta)= \pm {{\vec b}_{\vec k}}{}^{\pm} \frac{\partial
\beta_{\vec k}}{\partial\eta}^\pm,
$
where ${\hat e}_1$, ${\hat e}_2$ are unit vectors in the plane
perpendicular to ${\vec k}$ such that $({\hat e}_1,{\hat
e}_2,{\hat k})$ is a right-handed system. The function
$\beta_{\vec k}(\eta)$ obeys the following equation,
\begin{equation}
\label{evolution} \frac{\partial^2  \beta_k }{\partial
\eta^2}^\pm+\sigma \frac{\partial \beta_k}{\partial \eta}^\pm +
\left( k^2 \pm \lambda \frac{d \phi}{d \eta} k \right)
\beta_k^\pm(\eta) = 0.
\end{equation}

Some qualitative behaviour of the solutions of
eq.(\ref{evolution}), can be inferred from the simple case of a
constant $\frac{d \phi}{d \eta}$. Then the solutions are simply
linear superpositions of two exponentials. Only if $\lambda\left|
\frac{d \phi}{d \eta} \right|> k$, one of the two modes ($\pm$) is
exponentially growing. Otherwise both of them are either
oscillating or damped, as in ordinary magnetohydrodynamics. To
obtain significant amplification, coherent scalar field velocities
$\frac{d \phi}{d\eta}$ over a  duration are necessary, larger
velocities leading to larger amplification. The amplified mode is
determined by the sign of $\frac{d \phi}{d \eta}$. Modes with
wavenumber $k_{max}=\frac{1}{2} \lambda \left|\frac{d \phi}{d
\eta}\right|$ get maximally amplified.

For values of the scalar mass in the TeV range and temperature
above $100$ GeV, the cosmic friction term $ 2 a H \frac{d \phi}{d
\eta}$ in eq.(\ref{evolution}) is negligible compared to the mass
term. The scalar field oscillates and its velocity changes sign
periodically over a time scale much shorter than the Hubble time
at the epoch, so that both modes can be amplified. Each mode is
amplified during one part of the cycle and damped during the other
part of the cycle. Net amplification results when amplification
overcomes damping. It occurs for a limited range of Fourier modes,
peaked around  $k/m \sim \frac{1}{2}\lambda f$. The modes of the
EM fields are oscillating with (sometimes complicated) periodic
time dependence and an exponentially growing amplitude. Total
amplification is exponential in the number of cycles. For the
range of parameters in which fields are amplified, the amount of
amplification per cycle for each of the two modes is very well
approximated by the same  constant $\Gamma(k/m,\lambda f,\sigma)$.
In Fig. 1  we show: a) an example of the time dependence for a
specific mode and a selected set of parameters, and b) 
amplification rates as a function of the wave number $k$.

Another interesting approximate solution can be obtained for $m\ll
T$. In this extreme limit, it would take a time interval of the
order of the characteristic cosmic expansion time for the scalar
velocity to change its value significatively. We say that the
scalar rolls. Eq.(\ref{evolution})
can be approximated by a first order equation
$\sigma\frac{\partial\beta}{\partial \eta}^\pm + \left( k^2 \pm
\lambda \frac{d \phi}{d \eta} k\right) \beta^\pm(\eta) = 0$, which
can be solved exactly. The amplified mode is determined by the
sign of $\Delta \phi(\eta)= \phi(\eta)-\phi(0)$. Looking at
$\eta\sim \eta_{EW}$ the amplification is maximal for
$k_{max,EW}\sim 10^{-7} T_{EW}$.

A detailed discussion of the end of oscillations or rolling is
beyond the scope of this paper. However, we do know that once the
oscillations or rolling stop, the fields are no longer amplified
and obey a diffusion equation. Modes with wave number below the
diffusion value $k <k_\sigma T\sim 10^{-8} T$, where
$\frac{k_{\sigma}^2}{\sigma}\frac{1}{\eta_{EW}}=1$, remain almost
constant until the EW transition, their amplitude goes down as
$T^2$, and energy density as $T^4$, maintaining a constant ratio
with the environment radiation. Modes with $k/T>k_\sigma$ decay
quickly, washing out the results of amplification. We have seen
that for oscillating fields the range of momenta $k^{-1}>r_{Debye}$ 
that get amplified is not too different than $T$,
therefore scalar field oscillations have to occur just before, or
during the EW transition. In that case, the amplified fields do
not have enough time to be damped by diffusion. If the field is
rolling, momenta $k\ll T$ can be amplified, and therefore the
rolling can end sometime before the EW transition.

\begin{figure}
\vspace{-.1in} \centerline{
\psfig{figure=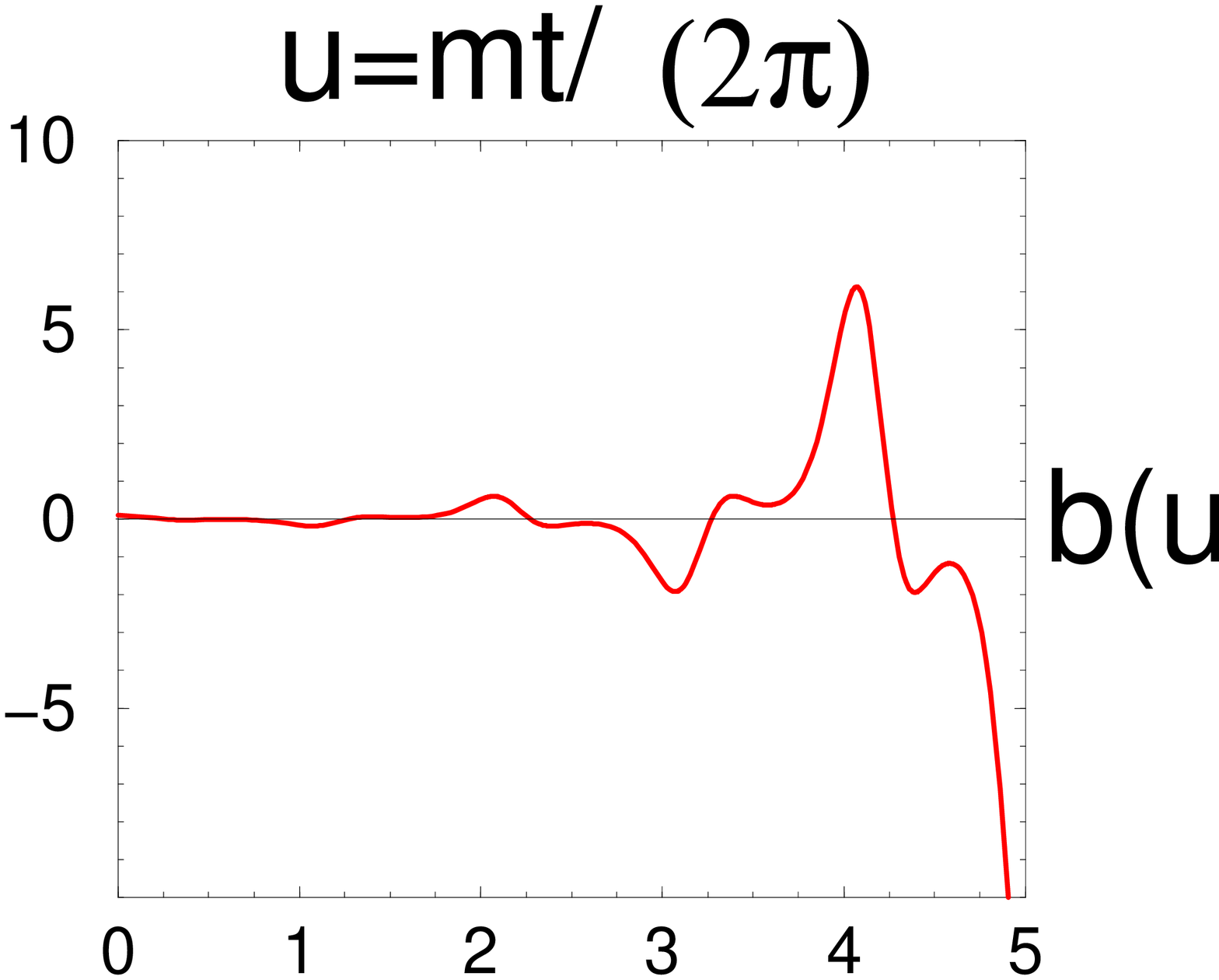,height=1.0in,height=1.5in}\hspace{.4in}
\psfig{figure=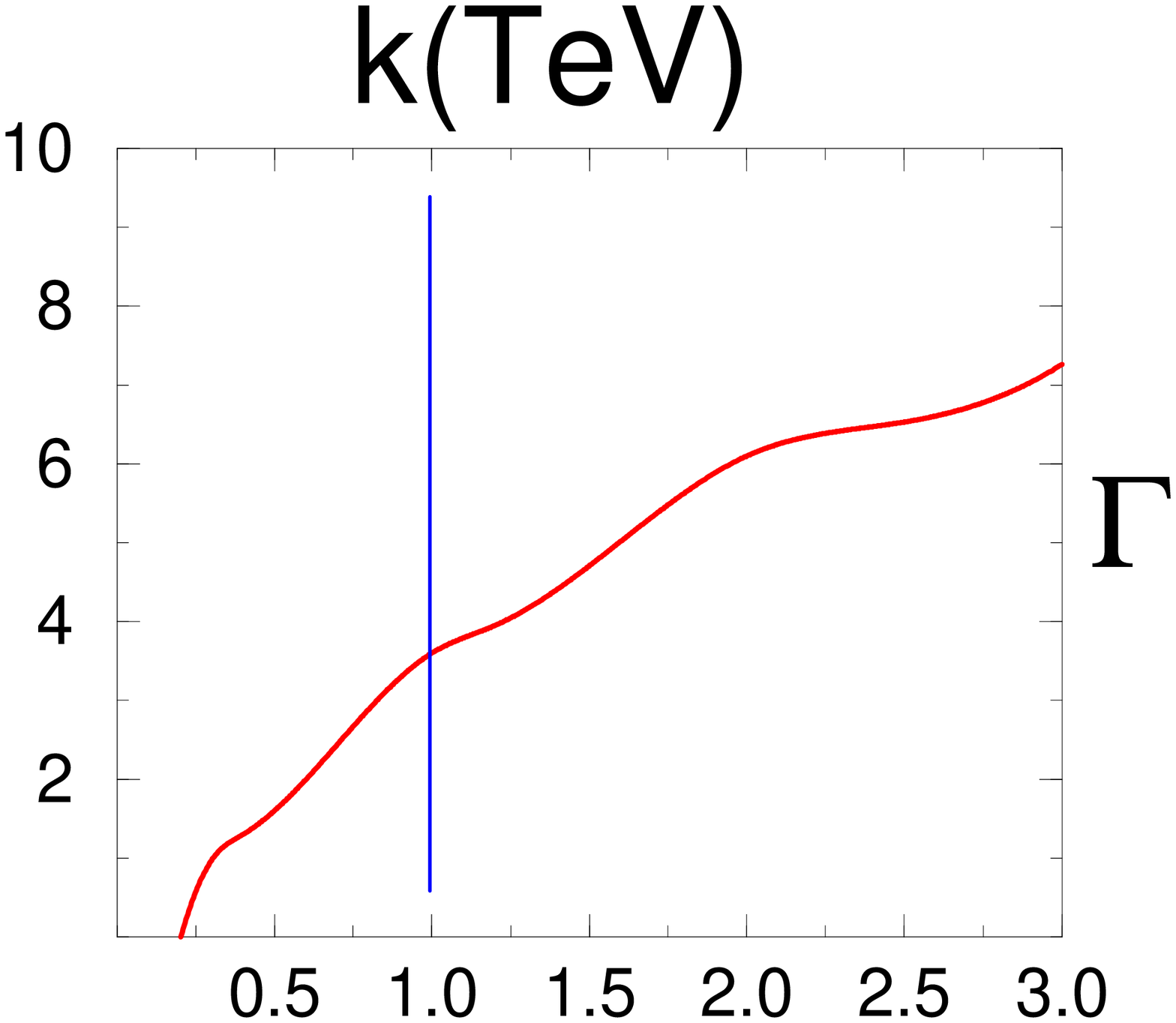,height=1.0in,height=1.5in} }
\caption{\label{field} { left: Time dependence for a selected mode,
$k=0.5 T$ with $m=4$ TeV, $\sigma=5 T$, $T=1$ TeV and $\Lambda=30$. The
function is periodic with exponentially growing amplitude; right:
amplification rate $\Gamma$ as a function of the wave number $k$
for the selected set of parameters. Total amplification is
exponential in the number of cycles ${\it n}$: ${\cal A}\equiv
e^{\Gamma {\it n}}$. To the right of the vertical line 
$k^{-1}<r_{Debye}$ and our assumption of a neutral plasma is not
valid.}}
\end{figure}

\section{Implications for Electroweak Baryogenesis}

To obtain the average magnetic energy density in amplified fields
we have to average over the initial conditions, which may result,
for example, from initial thermal or quantum fluctuations. We
assume translation and rotation invariance,
$
\langle B^{\pm}({\vec k}, \eta) {B^{\pm}({\vec \ell}, \eta)}^*
\rangle_{|\eta=0}= \delta^3({\vec k}-{\vec \ell})\ \delta_{+ -}
f^\pm(k).
$
Using the definition for magnetic power spectrum
$
P_B^\pm(k)=4\pi  k^3 f^\pm(k),
$
the average energy density in amplified fields $\rho_B =\frac{1}{8
\pi} <B^2>$, can be expressed as
$
\rho_B= \frac{1}{8 \pi}  \int d\ln k$ $\Biggl\{ P_B^-(k) \left|
{{\cal A}^-}({k},\eta)\right|^2 +
 P_B^+(k)
 \left| {{\cal A}^+}({k},\eta)\right|^2
\Biggr\}$, 
where ${{\cal A}^\pm}({k},\eta)$ are the
amplification factors for each one of the two magnetic modes.

Since in our case $E_{ {\vec k}}^\pm(\eta)= \pm \frac{1}{k}
\partial_\eta B_{ {\vec k}}^\pm(\eta)$, and therefore
$
\langle E^{\pm}({\vec k}, \eta) {B^{\pm}({\vec \ell}, \eta)}^*
\rangle= \pm \frac{1}{2} \frac{1}{k} \partial_\eta \langle
B^{\pm}({\vec k}, \eta) {B^{\pm}({\vec \ell}, \eta)}^* \rangle,
$
the CS number density stored in the amplified magnetic modes
$
\Delta n_{CS}=- \frac{g'^2}{4 \pi^2}  \int\limits_0^\eta
d{\widetilde \eta} \langle E\cdot B\rangle $, where $g'$ is the
hypercharge gauge coupling, can be related to $\rho_B$
\be
n_{CS}= \frac{g'^2}{\pi}  \int d \ln k \frac{1}{k} \rho_B (k)
\frac{\gamma^B_{AS}+ \gamma^\phi_{AS} } {1+ \gamma^B_{AS}
\gamma^\phi_{AS} }, 
\ee
introducing the asymmetry parameters,
$
{ \gamma}^B_{AS}= \frac{P_B^-(k) - P_B^+(k)} {P_B^-(k) +
P_B^+(k)},$ 
\ and 
$
\gamma^\phi_{AS} = \frac{\left| {{\cal
A}^-}({k},\eta) \right|^2 - \left| {{\cal A}^+}({k},\eta)
\right|^2} {\left| {{\cal A}^-}({k},\eta)  \right|^2 + \left|
{{\cal A}^+}({k},\eta) \right|^2}.
$
 Further, we can relate the fractional
energy density in coherent magnetic field configurations
$\Omega_B(k)=\rho_B(k)/\rho_c$ to the CS fractional number density
$n_{CS}/s$, assuming the universe is radiation dominated and the
amplification factors ${{\cal A}^\pm}_{k}$ are, as we have seen,
sharp functions of $k$, peaked at $k_{max}$\footnote{As we have
pointed out our description is valid for short wavenumber modes
$k<k_D \equiv r_D^{-1}$. $k_{max}$ is understood to be the
maximally amplified mode in the range $0<k<k_D$.},
\be
\frac{n_{CS}}{s}\simeq 0.01
 \frac{T}{k_{max}} \Omega_B(k_{max})  \frac{\gamma^B_{AS}+
\gamma^\phi_{AS} } {1+ \gamma^B_{AS} \gamma^\phi_{AS} } (k_{max}).
\ee

This Chern-Simons number will be released in the form of fermions
which will not be erased if the EW transition is strongly first
order\cite{Shaposhnikov:1986jp}, and will generate a baryon
asymmetry \cite{Giovannini:1998eg}, $\frac{n_{B}}{s}=-\frac{3}{2}
\frac{n_{CS}}{s}$.  An
equal lepton number would also be generated by the same mechanism
so that $B-L$ is conserved. Note that the fact that baryon number
asymmetry is generated at $k_{max}\ne 0$ does not mean that baryon
density is actually inhomogeneous on this short length scale
$L_{max} \sim 1/k_{max}$. Comoving neutron diffusion distance at
the beginning of nucleosynthesis is much longer than $L_{max}$, so
that by that time inhomogeneities would have been erased by free
streaming \cite{Jedamzik:1994ut}.
If $T/ k_{max}$ is not too different than unity, as we have seen
for the case of oscillating field, and $\gamma^B_{AS}$ and
$\gamma^\phi_{AS}$ are small, it is possible to obtain
$\frac{n_{B}}{s}\sim 10^{-10}$ and have strong magnetic fields
$\Omega_B\sim 1$ present during the EW transition. If $T/ k_{max}$
is large and $\gamma^\phi_{AS}$ is order unity as we have seen in
the rolling case,  it is not possible to have strong magnetic
fields without producing too many baryons.

\section{Experimental Signatures of HyperAxions in Colliders}

A pseudoscalar field that couples to hypercharge topological
number density as we have described, could be produced and detected in
colliders \cite{Brustein:1999it}. Since the hypercharge photon is
a linear combination of the ordinary photon and the $Z$, the
hypothetical particle could be produced, in hadronic or leptonic
colliders, in association with photons or $Z$'s through the
channels $f \overline{f} \rightarrow Z^{*},\gamma^{*} \rightarrow
Z X$ and $f\overline{f} \rightarrow Z^{*},\gamma^{*}\rightarrow
\gamma X$, (here $Z^*$ and $\gamma^*$ denote a virtual $Z$ and a
virtual photon, and $f$ is a charged fermion). The produced
particle would decay into two neutral gauge bosons, $\gamma
\gamma$, $\gamma Z$ or $Z Z$. The experimental signature of these
processes would , then, be a triplet of neutral gauge bosons
produced in well defined ratios. The almost isotropic angular
distribution of momenta of the outgoing bosons could help in
separating it from the QED background, that is strongly peaked in
the forward/backward directions.

In the simple case of a singlet elementary pseudoscalar whose only
coupling to SM fields is to hypercharge topological density, the
extra field can be described by two parameters with units of mass:
the mass, $m$, and the inverse coupling, $M=1/\lambda$. The plots
in Fig. 2  show our estimation of the regions of the $(m,M)$
parameter space where the particle signature could be separated
from the background, and so the particle can be detected in
present or future colliders. Below the curves, a sufficient number
of events to allow detection are expected. Since the particle has
not been detected yet, the first plot, that corresponds to LEPII
and RunI of the Tevatron, can be used to rule out an interesting
region of parameters. In future colliders detection capabilities
will be increased significatively in the range of parameters relevant
for baryogenesis, as shown in Fig 2.

\begin{figure}
\vspace{-.1in}
\centerline{\psfig{figure=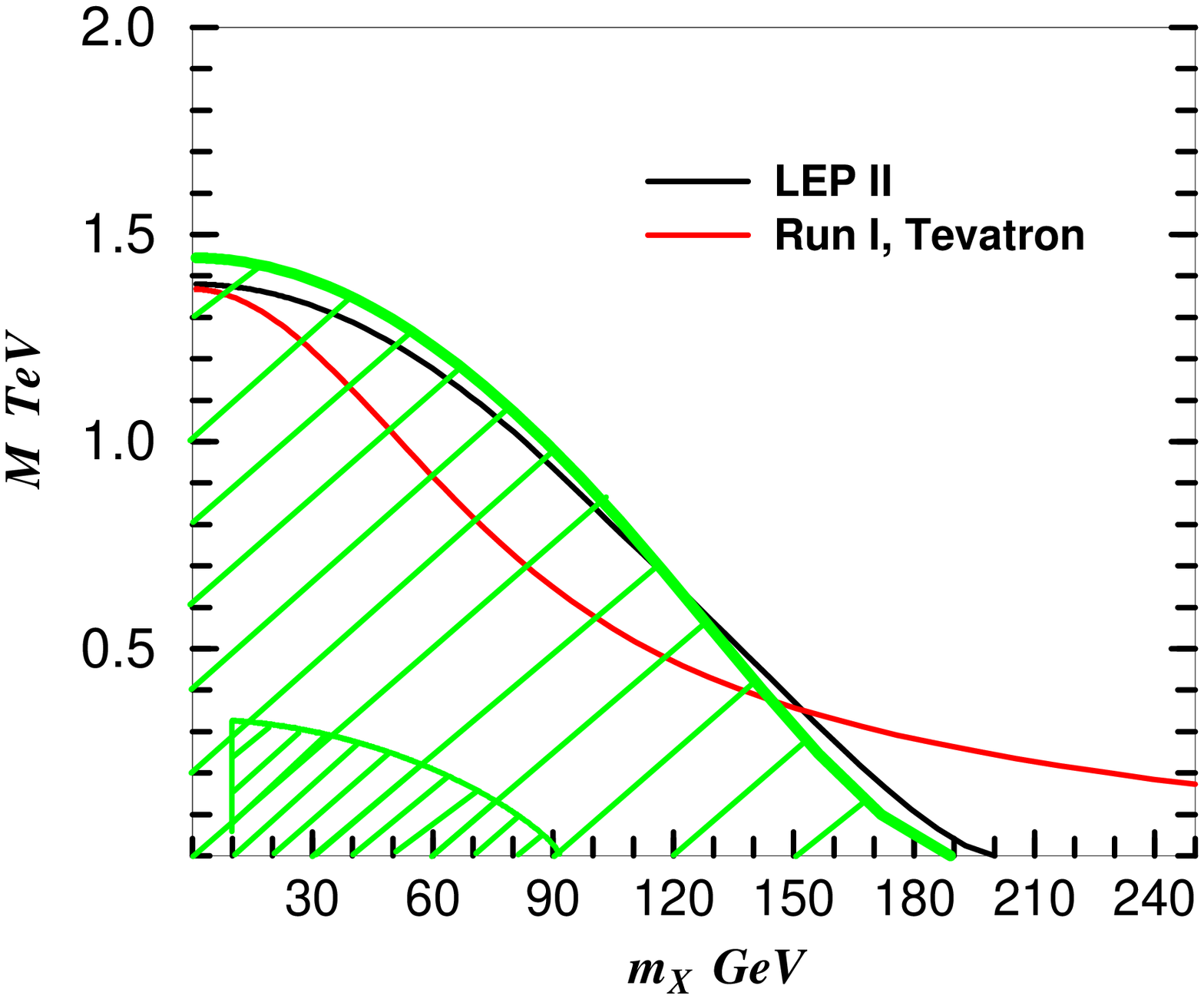,height=1.9in}\hspace{.3in}
\psfig{figure=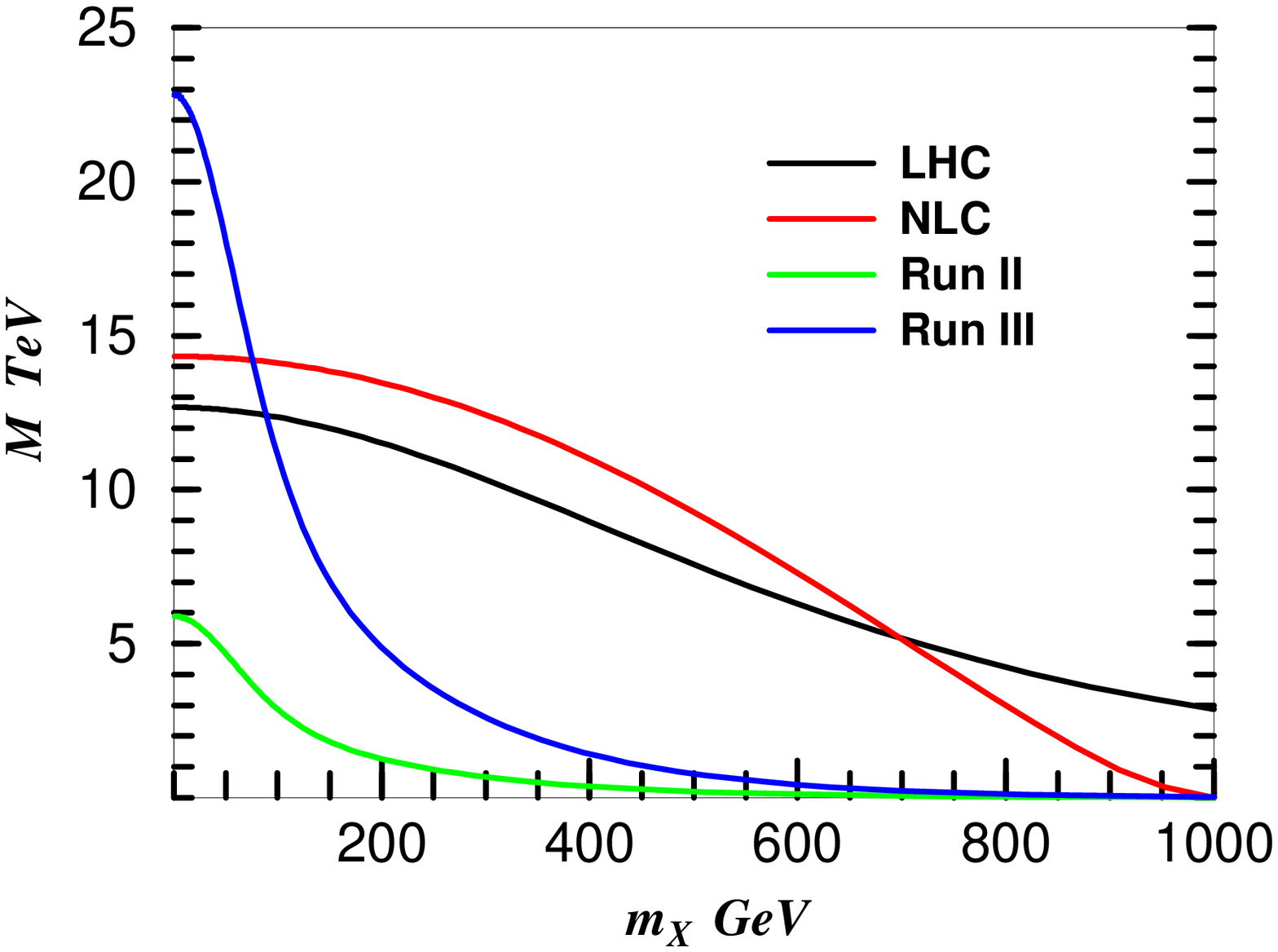,height=1.9in}}
\caption{\label{leptev} {Expected number of HCA's in present and
future colliders. The curves correspond to 10 expected events
required for the signal to be distinguished from the SM
background. The shadowed region is experimentally ruled out.} }
\end{figure}

\end{document}